Coexistence of Superconductivity and Magnetism in $MoSr_2RCu_2O_8$ (R=rare-earth, Mo-1212)

I. Felner, E. Galstyan, I. Asulin, A. Sharoni, and O. Millo

Racah Institute of Physics, The Hebrew University, Jerusalem, 91904, Israel.

The properties of the $MoSr_2RCu_2O_8$ (R=rare earth) system are found to systematically change with the contraction of the R ions. For the light R ions (La-Nd) the samples are paramagnetic down to 5 K, whereas in the intermediate range (Sm-Tb), the Mo sublattice orders antiferromagnetically at $T_N$, ranging from 11 to 24 K. For the heavy R ions, Ho-Tm and Y, superconductivity appears at $T_C$ in the range 19-27 K and antiferromagnetism sets in at $T_N < T_C$. This latter behavior resembles most of the magneto-superconductors, but is in sharp contrast to the iso-structural $RuSr_2RCu_2O_8$ system where $T_N > T_C$.

## Introduction

Superconductivity (SC) and magnetism are two different ordered states into which substances can condense at low temperatures and in general these states are inimical to one another. It has generally been believed that the conduction electrons in a metal cannot be both ferromagnetically (FM) ordered and superconducting. In conventional s-wave superconductors, local magnetic moments break up the spin singlet Cooper pairs, and hence strongly suppress SC, an effect known as pair-breaking. Therefore, a level of magnetic impurity of only 1 %, can result in a complete loss of SC. In a limited class of *intermetallic* systems, SC occurs even though magnetic ions with a local moment occupy all of one specific crystallographic site, which is well isolated and decoupled from the conduction path. The study of this class of magnetic-superconductors was initiated by the discovery of the $RRh_4B_4$ and $RMo_6S_8$ compounds (R=rare-earth)[1], and has been revitalized by the discovery of the $RNi_2B_2C$ system[2]. In all three systems, the SC and antiferromagnetic (AFM) states coexist and $T_N < T_C$ (except for $DyNi_2B_2C$). On the other hand in the $CeRh_{1-x}Co_xIn_5$ system, $T_N$ (~ 3-4) K is higher than $T_C$ (~ 1-2 K)[3]. More recently, coexistence of SC ($T_C$~ 0.3 K) and FM was found in URhGe ($T_M$=9.5 K)[4] and in high-quality crystals of the itinerant-ferromagnet $ZrZn_2$ ($T_M$ = 29 K)[5]. In all systems mentioned, the SC state can be explained within the context of the conventional Bardeen-Cooper-Schriefer (BCS) theory.

Coexistence of SC and magnetism was discovered a few years ago in $RuSr_2R_{2-x}Ce_xCu_2O_{10}$ (R=Eu and Gd, Ru-1222)[6-7], and more recently in $RuSr_2GdCu_2O_8$ (Ru-1212)[8]. Both Ru-based layered cuprate systems evolve from the $YBa_2Cu_3O_x$ structure where the Ru ions reside in the Cu (1) site, and only one distinct Cu site (corresponding to Cu (2) in $YBa_2Cu_3O_x$) with fivefold pyramidal coordination, exists. The SC charge carriers originate from the $CuO_2$ planes and the magnetic state is confined to the Ru layers. The Ru-1222 materials display a weak-ferromagnetic transition at $T_M$= 125-180 K and bulk SC below $T_C$ = 32-50 K ($T_M > T_c$) depending on the R/Ce ratio and/or oxygen concentration[9]. The magnetic state of the Ru sublattice, which coexists with the SC state, is not affected by the presence or absence the SC State, indicating that the two states are practically decoupled[10]. Ru-1212 is SC around 32-35 K and orders magnetically at 135 K. Its magnetic properties are still not understood. Magnetization studies show definitely a FM transition, whereas neutron diffraction measurements indicate that the Ru moments are AFM ordered. Recently we have noticed that the Ru ions can be replaced completely by Mo ions and that the $MoSr_2R_{1.5}Ce_{0.5}Cu_2O_{10}$ (Mo-1222) and $MoSr_2R_1Cu_2O_8$ (Mo-1212) systems can be obtained with most of the R elements (Pr-Yb and Y) as nearly single-phase materials. In a recent publication[11] we have shown that, in Mo-1222, the SC and the magnetic states compete with each other. Materials, which are SC, are not AFM, and vice versa. This is in contrast to the Ru-1222 system mentioned above.

In the new Mo-1212R system presented here the magnetic and/or the SC states are determined by the ionic radii of R. The light R samples (Pr and Nd) are PM down to 4 K, the middle R samples (Sm, Gd and Tb) are AFM ordered. Surprisingly, in the heavy R samples (Ho, Er, Tm and Y) coexistence of SC and AFM is observed. STM measurements on Mo-1212Y show that the surface contains SC regions with gaps of 6-7 meV. In contrast to Ru-1212 and Ru-1222 systems, in Mo-1212R $T_C$ (19-27 K) is higher than $T_N$. In that sense, this system behaves similarly to most of the conventional intermetallic systems discussed above.

## Experimental Details

Ceramic samples with nominal composition $MoSr_2RCu_2O_8$ (Mo-1212R) were prepared by a solid-state reaction technique. Prescribed amounts of $R_2O_3$, $SrCO_3$, Mo, and CuO were mixed and pressed into pellets and (a) preheated at 750° C for 1 day. (b) The products were cooled, reground and sintered at 1000-1030° C for 2 days under an oxygen atmosphere and then furnace cooled. For $MoSr_2YCu_2O_8$ sample (i), step (b) was done in air, whereas sample (iii) was heated at 950° C under nitrogen prior to step (b). Powder X-ray diffraction (XRD) measurements indicate that all samples are nearly single-phase (~96%) materials and confirmed the tetragonal structure (S.G. P4/mmm). The XRD patterns left a few minor reflections, most of them belonging to $R_2O_3$ and to the $SrMoO_4$ phases. All attempts to completely get rid of them were unsuccessful. For the R=Yb and Lu samples, the extra peak intensity exceed 25%. Zero-field-cooled (ZFC) and field-cooled (FC) dc and ac magnetic measurements in the range of 2-400 K were performed, as described in Refs. 6 and 9. For the STM measurements the samples were mechanically polished with 0.25 μm diamond lapping compound. The tunneling dI-dV vs. V spectra were acquired at 4.2 K along with the topographic images, while momentarily disconnecting the feedback loop, with tunneling resistances between 100 MΩ to 1GΩ.

## Experimental details

Least squares fits of the XRD patterns of the $MoSr_2RCu_2O_8$ compounds on the basis of a tetragonal structure yield the *a* and *c* lattice parameters. The variation of *a* (shown in Fig. 1), is attributed to the lanthanide contraction of the $R^{3+}$ ions. The *c* constant can be considered as remaining constant as 11.50(2) Å for all compounds. For R=Lu several unidentified extra peaks are observed and its *a* value is not included in Fig. 1. The morphology of Mo-1212Y detected by scanning electron microscopy, shows a smooth and uniform surface, with typical grain size of 2-3 μm. We also observed several separate spherical grains of the Pauli-paramagnetic SrMoO$_4$ phase[12], which do not affect the magnetic properties as discussed below.

**PM in MoSr$_2$RCu$_2$O$_8$ (R=La, Pr and Nd).** The dc magnetic susceptibility [χ(T) (=M(T)/H)] curves of Mo-1212R (R=La ,Pr and Nd ), exhibit normal PM behavior down to 5 K and all isothermal M(H) (up to 5 T) curves are linear. The χ(T) curves can be fitted by the Curie-Weiss (CW) law: χ =χ$_0$ +C/(T-θ), where χ$_0$ is the temperature independent part of χ, C is the Curie constant, and θ is the CW temperature. The extracted effective paramagnetic moment (P$_{eff}$) for Mo-1212La is 0.94μ$_B$. the value of P$_{eff}$ = 3.1(1) and 2.1(1) μ$_B$ obtained for R=Pr and Nd, are lower than the 3.56 and 3.62 μ$_B$ values expected for Pr$^{3+}$ and Nd$^{3+}$. This reduction is probably due to strong crystal field effects, which exist in these materials similar to that observed in the Ru-1222 system[13]. All θ values obtained are negative: -12.2, –25(1) and -1.4(1) K respectively.

**AFM in MoSr$_2$RCu$_2$O$_8$ (R=Sm, Gd, Tb and Dy).**

The ZFC and FC curves of Mo-1212Sm measured at 50 Oe, is presented in Fig. 2. The two branches merge at T$_N$ =11 K, (our definition of T$_N$). The observed peak is related to the AFM order of the Mo sublattice. The peak does not shift with H. Similar behavior was observed for R= Tb and Lu (T$_N$ =23 and 22 K respectively). Fig. 2 shows (inset) that a small hysteresis loop open at 5 K. Above 0.8 T the M(H) curve is linear (up to 5 T). The. For R=Dy, neither irreversibility nor a peak are observed in the χ(T) curves down to 5 K, probably because the small AFM signal of Mo (0.001 emu/mol Oe for R=Sm ) is masked by the high PM susceptibility of Dy$^{3+}$. Above T$_N$, the χ(T) curve for R=Sm does not follow the CW law. The PM parameters for R=Tb and Dy, are: P$_{eff}$= 8.8 and 9.3 μ$_B$ (smaller than the expected values for R$^{3+}$ free ion R$^{3+}$) and θ =-15(1) and –5.5 K respectively.

The MoSr$_2$GdCu$_2$O$_8$ sample shows three magnetic transitions at 2.6, 24 K and at 284(2) K (Fig. 3). The peak at T$_N$(Gd)= 2.6(1) K is due to the AFM ordering of the Gd sublattice and it is very close to the T$_N$ values of Gd in GdBa$_2$Cu$_3$O$_x$ and Ru-1212[14]. The peak at 24 K (the highest for Mo=1212R, is probably the AFM ordering of the Mo sublattice. At the present moment we cannot explain the third transition at 284 K, which resembles the one observed at 184 K in Mo-1222[11]. The M(H) curves are linear and no hysteresis is observed. At 5 K, for H> 3 T, a tendency toward PM saturation (of Gd ions) is observed, and the moment 6.6 μ$_B$/Gd obtained at 5 T, is close to theoretical saturated Gd (7 μ$_B$) value. Alternatively, the two peaks at 284 and 24 K, may be associated with a minor fraction of Gd$_2$CuO$_4$ phase (which were observed only for $Gd_2CuO_4$) not detectable by XRD and are related to the magnetic transition and to the spin reorientation of the *Cu* moments respectively, due to a Gd-Cu interaction[15]. Mossbauer studies on $^{57}$Fe doped material have been performed at various temperature. Generally speaking, the obtained results support the existence of a minor fraction of $Gd_2CuO_4$, however, this interpretation is not exclusive and the first scenario obtained cannot excluded.

**Coexistence of SC and AFM in $MoSr_2RCu_2O_{10}$ (R=Ho, Er, Tm and Y).**

For the heavy R ions**,** in addition to the AFM discussed above, a SC state is induced and both states coexist. We shall start with the non-magnetic Y ions, which permit an easier direct interpretation of the intrinsic Mo magnetism. The magnetic study of Mo-1212Y was performed on three different samples as described above. Sample (i) (sintered in air), is not SC. The ZFC and FC curves are *positive,* show a peak at 11 K, and merge at 17 K. The magnetic features for sample (ii) (sintered under oxygen) are presented in Fig. 4. Note the different behavior of the ZFC and FC branches. The negative ZFC signal indicates clearly a SC state with an onset at $T_C$= 22(1) K, a value which was also confirmed by ac and resistivity measurements. On the other hand, a pronounced peak is observed in the FC branch at 11 K, and the two branches merge at $T_N$=17 K, as for sample (i), indicating an AFM ordering of the Mo sublattice, below $T_C$. Note the negative values in the FC curve around $T_N$. The ZFC and FC branches for sample (iii) (annealed first under nitrogen) are both negative below $T_C$= 22 K (not shown) and the FC branch shows a clear peak at 11 K. The shielding and Meissner fractions (without correcting for diamagnetism) exceed 20% and 16%, indicating clearly bulk properties of the SC state. With the purpose of acquiring information about the critical current density $J_C$, we have measured at 5 K the magnetic hysteresis (Fig. 5) of sample (iii). The inset shows the linear behavior at higher fields typical of an AFM ordering. Following Bean's approach $J_C(H)=30 \Delta M/d$, where $\Delta M$ is the difference in the M at the same H, and d=2.5 μm, we obtained: $J_C= 2 *10^5$ A/cm$^2$ (at H=0) a value, which compares well with $J_C$ obtained in Ru-1222 under the same conditions[16].

Above $T_N$, the χ(T) curve for Mo-1212Y follows the CW law. Note that: (1) the Y ions, (2) the $SrMoO_4$ phase and (3) the Cu contribution, which is roughly temperature independent (1.8-2*10$^{-4}$ emu/mol Oe)[11], do not contribute to C. Therefore, the $P_{eff}$ =1.78 and 1.89 μ$_B$ obtained for samples (ii) and (iii) respectively, correspond only to the Mo ions. These values are in good agreement with 1.73 μ$_B$ expected for Mo$^{5+}$ (4d$^1$, S=0.5). Therefore, we argue with high confidence, that the prominent AFM features shown in Figs. 2, 4-6, (as well as the peak at 24 K for Mo-1212Gd Fig. 3) are related to the Mo$^{5+}$ sublattice. The SC state is confined to Cu-O layers and similar to the ruthenates, the two states coexist on a microscopic scale.

Similar coexistence was also observed in the heavy R ions: Ho, Er and Tm. Fig. 6 exhibits the ZFC and FC curves for Mo-1212Er measured at 30 Oe. One definitely sees the irreversibility below $T_N$=24(1) K and the peaks in both branches. The SC transition (above $T_N$) is masked by the high PM contribution of $Er^{3+}$, and onset of the $T_C$= 27(1) K was obtained from the real part of the ac susceptibility $`\chi(T)$ shown in the inset. $`\chi(T)$ was measured under two external fields. The merging temperature ($T_P$) of two curves is not at $T_N$. Since the Mo-1212R materials are all ceramic granular materials, the broad SC transition occurs via two stages. At $T_C$ (which is not severely affected by H) the grains become SC, whereas $T_P$ is the weak-Josephson inter-grain transition, which is affected dramatically by H.

Similar behavior was observed for R=Ho and Tm, where $T_N$ =16 and 17 K and $T_C$ (deduced from $`\chi$) of 19 and 24 K were obtained. The PM parameters obtained are: $P_{eff}$ =8.5, 9.4 and 5.7 $\mu_B$ and $\theta$ = -6, -17 and –7 K for R= Ho, Er and Tm respectively. These $P_{eff}$ moments (even without subtracting the Mo contribution) are lower than the expected values for free $R^{3+}$ ions probably due to strong crystal field effects as mentioned above. The negative $\theta$ values obtained are consistent with an AFM order. For R=Yb, only a SC transition at $T_C$= 28 K is observed (Fig. 1). However the extra phases observed in the XRD pattern, do not permit determination of the composition of the tetragonal phase.

As a final point of interest, we discuss the STM measurements performed at 4.2 K on Mo-1212Y sample (ii). The STM topographic images reveal granular surface morphology, with rms roughness amplitude around 5 nm and surface features of typical 50 to 100 nm lateral size. Some of the surface grains revealed gapless (nearly Ohmic) tunneling spectra, whereas for many clear SC gaps were observed in the tunneling spectra (Fig. 4, inset). The gap width exhibited spatial variations in the range 6-7 meV, thus the ratio $2\Delta/k_BT_c$ is around 6.8, within the range observed for various cuprate superconductors.[6,17] We note that the gaps vanished at $T_C$ and were independent of the STM voltage and current setting. This rules out the possibility that they are due (even partly) to the Coulomb blockade,[18] and therefore they are unambiguously associated only with the SC state.

**In conclusion**, we demonstrate that in Mo-1212 ($Mo^{5+}$), the physical state depends strongly on the ionic radii of the R ions. For large R (La-Nd) ions, the samples are PM (Fig. 1). Once the R (Sm-Tb) ionic radii are contracted, AFM order (at $T_N$ ranging from 11-24 K) of the Mo sublattice is induced. Further contraction (R=Ho-Tm and Y) induces a SC state in the Cu-O layers, which coexist, with the AFM state through effectively decoupled subsystems. Samples prepared in air are AFM only, and SC can be achieved upon appropriate annealing.

This rather surprising results raises a question as to why Mo-1212 ($T_C>T_N$) behave so differently than the Ru-1212 system in which $T_C<<T_N$.

## Acknowledgments

We are grateful to Prof. S. Reich and to L. Polachek for their assistance. This research was supported by the Israel Academy of Science and Technology and by the Klachky Foundation for Superconductivity.

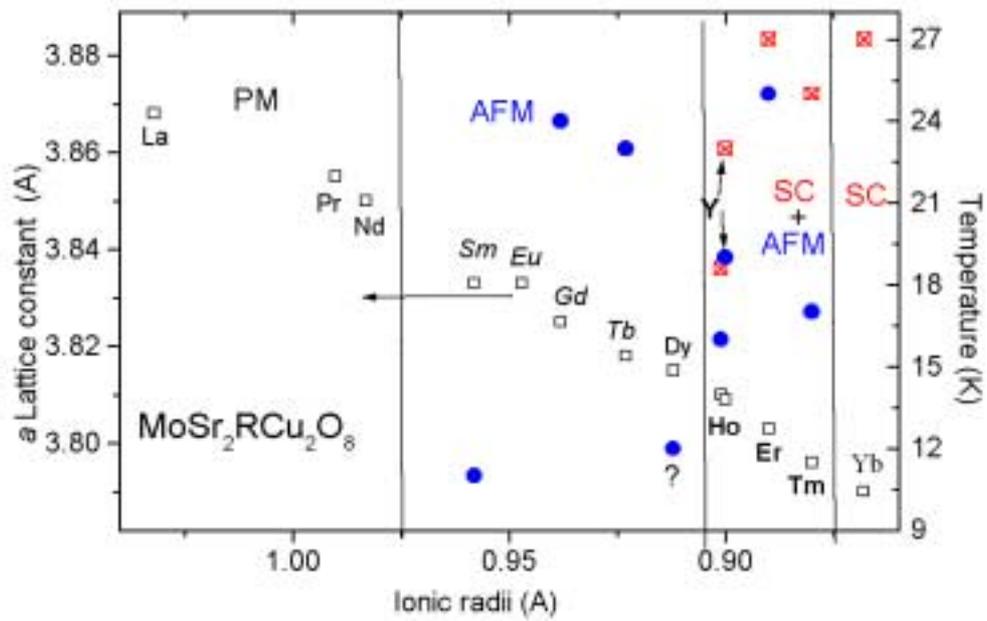

Fig. 1 The AFM and SC phase diagram (including the $T_C$ and $T_N$ values) and the $a$ lattice parameter as a function of the ionic radii in the $MoSr_2RCu_2O_8$ system.

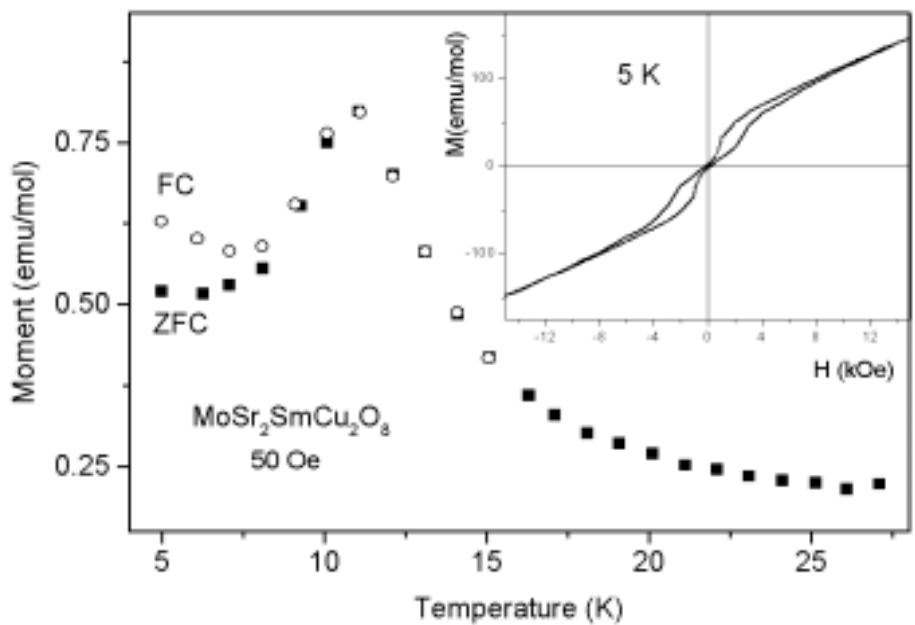

Fig. 2. ZFC and FC magnetic curves and the isotherm magnetization curve at 5 K of $MoSr_2SmCu_2O_8$

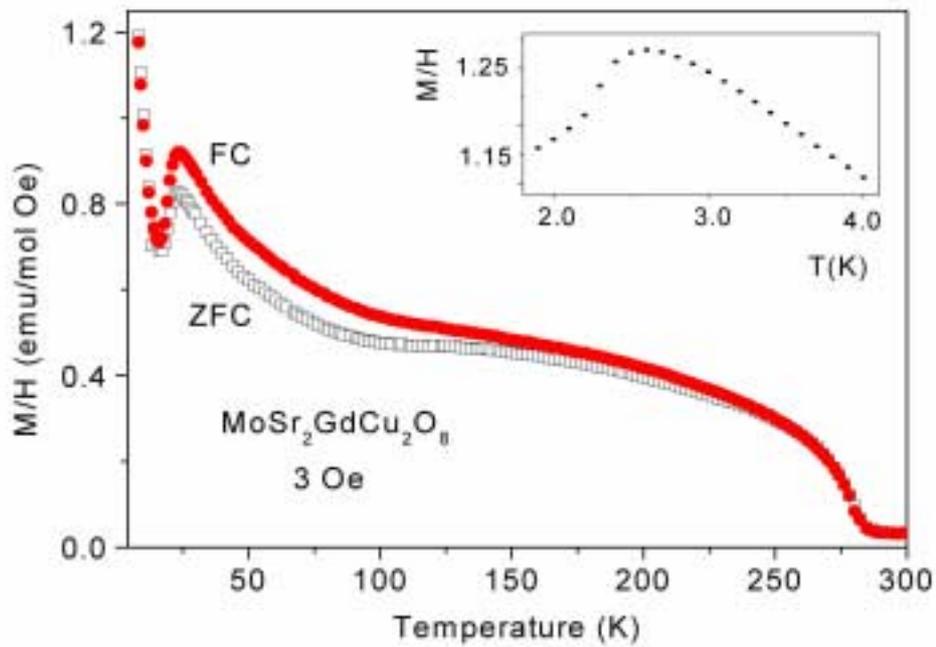

Fig. 3. ZFC and FC susceptibility curves for MoSr$_2$GdCu$_2$O$_8$. The inset shows the magnetic transition at 2.6 K of the Gd sublattice.

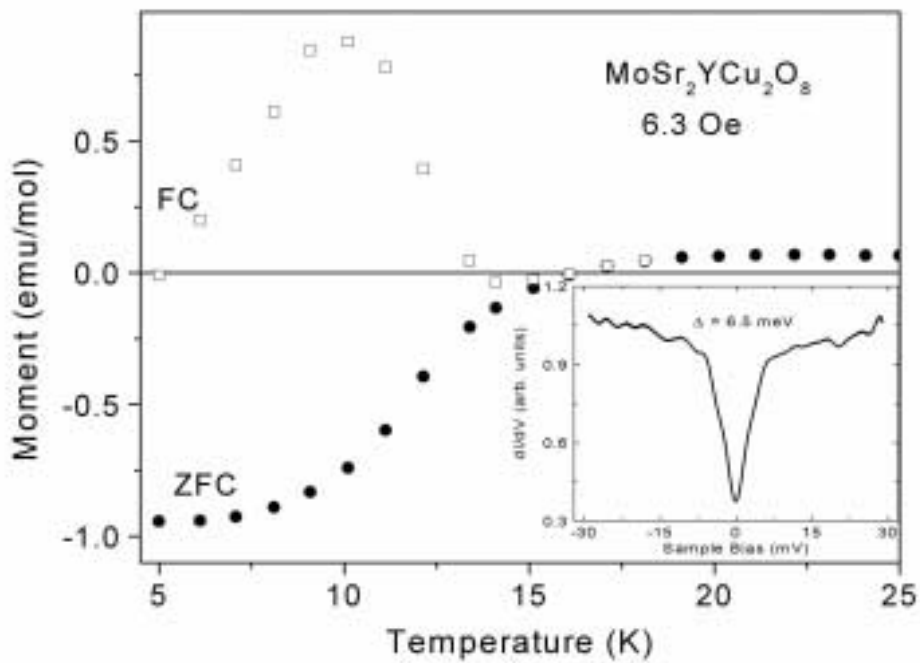

Fig. 4. ZFC and FC magnetic curves and a typical tunneling spectrum acquired at 4.2 K of MoSr$_2$YCu$_2$O$_8$ (sample (ii)). Note the negative values of the FC branch.

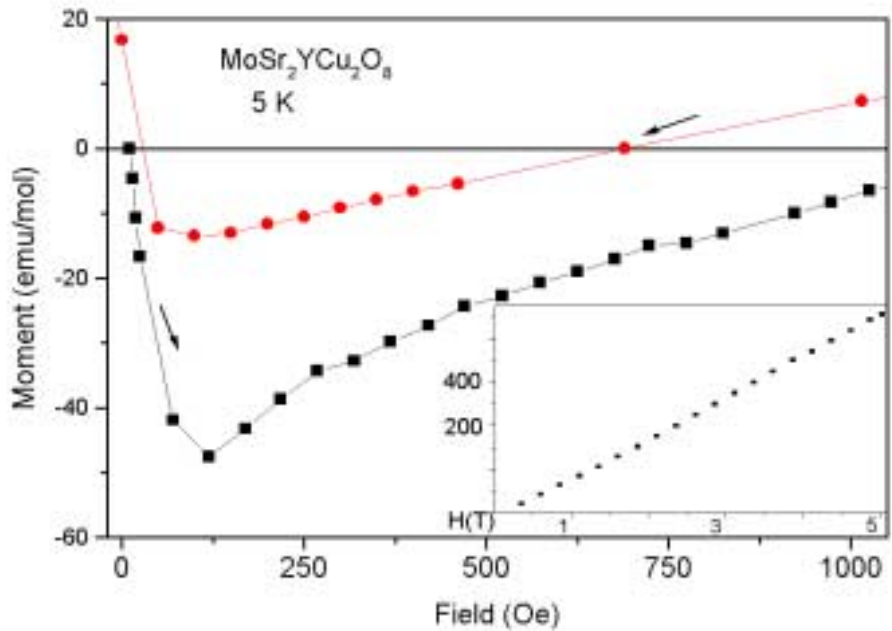

Fig. 5. The magnetization (inset) and the low field hysteresis loop at 5 K of $MoSr_2YCu_2O_8$ (sample (iii)

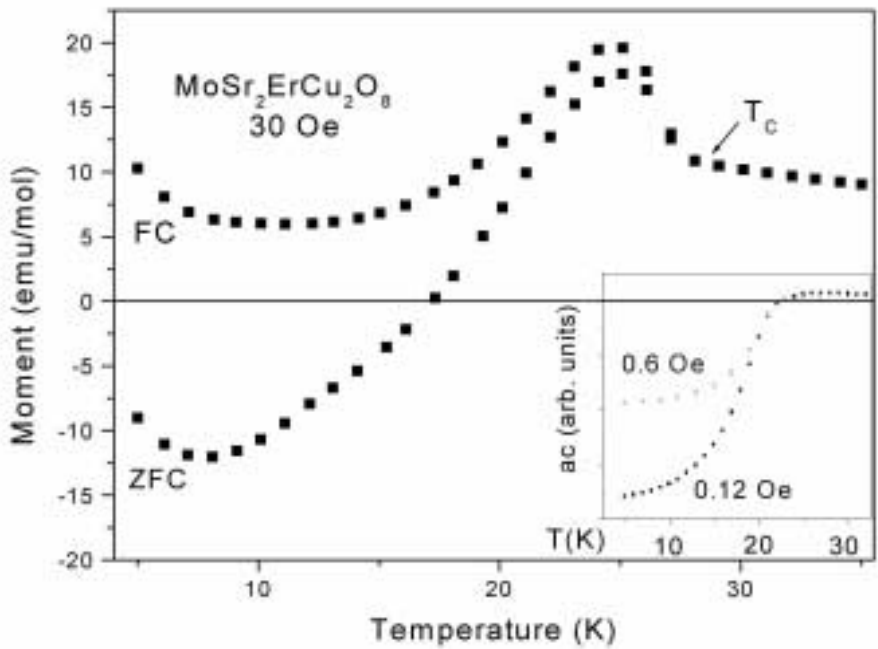

Fig. 6 dc ZFC and FC curves ac susceptibility curves of $MoSr_2ErCu_2O_8$